\documentclass[titlepage,12pt,a4paper,reqno]{article}
\usepackage{amsmath,amssymb,amsfonts,graphics}
\usepackage[linktocpage=true,colorlinks,pdftex]{hyperref}
\usepackage{xcolor}
\colorlet{linkequation}{blue}
\usepackage{bm}
\usepackage{doi}
\usepackage{hyperref}
\hypersetup{colorlinks, citecolor=violet, filecolor=black, linkcolor=black, urlcolor=blue}
\newcommand*{\refeq}[1]{%
  \begingroup
    \hypersetup{
      linkcolor=linkequation,
      linkbordercolor=linkequation,
    }%
    \ref{#1}%
  \endgroup
}
\allowdisplaybreaks

\begin{document} 


\begin{titlepage}

\centerline{\LARGE \bf On the existence of the field line solutions} 
\medskip
\centerline{\LARGE \bf of the Einstein-Maxwell equations}
\vskip 1.5cm
\centerline{ \bf Ion V. Vancea }
\vskip 0.5cm
\centerline{\sl Grupo de F{\'{\i}}sica Te\'{o}rica e Matem\'{a}tica F\'{\i}sica, Departamento de F\'{\i}sica}
\centerline{\sl Universidade Federal Rural do Rio de Janeiro}
\centerline{\sl Cx. Postal 23851, BR 465 Km 7, 23890-000 Serop\'{e}dica - RJ,
Brasil}
\centerline{
\texttt{\small ionvancea@ufrrj.br} 
}

\vspace{0.5cm}

\centerline{14 Novembro 2017}

\vskip 1.4cm
\centerline{\large\bf Abstract}

The main result of this paper is the proof that there are local electric and magnetic field configurations expressed in terms of field lines on an arbitrary hyperbolic manifold. This electromagnetic field is described by (dual) solutions of the Maxwell's equations of the Einstein-Maxwell theory. These solutions have the following important properties: i) they are general, in the sense that the knot solutions are particular cases of them and ii) they reduce to the electromagnetic fields in the field line representation in the flat space-time. Also, we discuss briefly the real representation of these  electromagnetic configurations and write down the corresponding Einstein equations.

\vskip 0.7cm 

\noindent

\noindent 

{\bf DOI} \doi{10.1142/S0219887818500548 }

\end{titlepage}


\section{Introduction}

Recently, there has been an increasing interest in the 
topologically non-trivial configurations of the electromagnetic field in the flat space-time \cite{Trautman:1977im,Ranada:1989wc,Ranada:1990,Ranada:1992hw}. These fields are associated to the solutions of the Maxwell's equations without sources subject either to the null field or more general conditions \cite{Bateman:1915}. The most known class of field line solutions of the Maxwell's equations have the knot topology given in terms of complex scalar Hopf maps $S^3 \rightarrow S^2$ over the compactified space-like directions. The basic physical and geometrical properties of the field line solutions were investigated in a series of early papers \cite{Ranada:1992hw}-\cite{Ranada:1997}. More recent studies have focused on other features of the electromagnetic fields in terms of field lines: in \cite{Irvine:2008,Irvine:2010} the authors studied the connection between the knotted and linked solutions, the dynamics of the electric charges in this background and of the knotted fields were investigated in \cite{Kleckner:2013}-\cite{Ranada:2017ore} and the topological quantization was discussed in \cite{Ranada:1998vp}-\cite{Arrayas:2012eja}. The generalization of the above solutions to the torus knot topology was given in \cite{Arrayas:2011ia}-\cite{Hoyos:2015bxa} and field lines and Hopf solutions were constructed in the non-linear electromagnetism in \cite{Goulart:2016orx}-\cite{Alves:2017zjt}. The importance of the  electromagnetic fields in terms of field lines is emphasised by a large range of phenomena in which they seem to be present: in fluid physics \cite{Alves:2017ggb,Alves:2017zjt}, atmospheric physics \cite{Ranada:1996}, liquid crystals \cite{Irvine:2014}, plasma physics \cite{Smiet:2015}, optical vortices \cite{Ren:2008zzf,deKlerk:2017qvq} and superconductivity
\cite{Trueba:2008sc}. (For a review of the subject and some of its applications see \cite{Arrayas:2017sfq} and the references therein).

The existence of the general local field line solutions is a fundamental step in constructing particular field configurations, for example Hopf and torus knot fields, since it is the field lines rather than the electromagnetic potentials that contain the information about the topology of a given field configuration \cite{Ranada:1989wc,Ranada:1990}. In this bottom to top approach, the connection between the field lines and the knot topology is given by the complex scalar maps that characterize the integral lines of the electric and magnetic components. Therefore, by proving the existence of the field line solutions to the Maxwell's equations one is able to describe a larger class of electromagnetic field configurations that contains the topological fields as a subclass specified by a certain number of parameters.

Our goal in this paper is to prove the existence of a general class of local field line solutions to the Maxwell's equations without sources on an arbitrary hyperbolic manifold in the Einstein-Maxwell theory. These solutions are the natural generalization of the flat space-time field configurations from \cite{Ranada:1989wc,Ranada:1990} to the gravitating electromagnetic field. They are parametrized by two locally smooth complex scalar fields $f$ and $g$ which satisfy a set of constraints that result from the representation of both electric and magnetic fields in terms of their own field lines\footnote{ The scalars $f$ and $g$ are geometrical fields, not physical. They describe the equations of the field lines. Therefore, there is no question about them carrying electric charges and coupling with the electromagnetic field.}. The local solutions are the most general ones on a curved space-time because their global continuation depends on the topological and geometrical properties of the underlying manifold. A particular solution was obtained in \cite{Kopinski:2017nvp} for the case of the cylinder topology that can be obtained from our local analysis in the case of the $\mathbb{R} \times S^3$ space-time manifold. Other attempts to generalize the Hopf knots to the gravito-electromagnetic fields were performed in \cite{Dalhuisen:2012zz,Swearngin:2013sks,Thompson:2014pta,Thompson:2014owa} in the context of higher spin fields.

The paper is organized as follows. In Section 2 we establish our notations following mainly \cite{Gourgoulhon:2007ue} and review the Einstein-Maxwell equations in the $1+3$ formalism that is best suited for addressing the existence of the field line solutions locally on the hyperbolic space-time manifolds. In Section 3 we propose a new expression for the electric and magnetic fields in terms of the field lines. Our proposal generalizes naturally the field configurations given in \cite{Ranada:1989wc,Ranada:1990} in the Minkowski space-time. Then we prove by direct calculations that the new fields satisfy the local Maxwell's equations. Although the computational steps follow closely the ones done in the flat space-time, the calculations are non-trivial due to the presence of the gravitational field. As in the Minkowski space-time, the electric and magnetic fields are orthogonal to each other and are parametrized by two arbitrary complex scalar fields $f$ and $g$ associated to the field lines. We show that $f$ and $g$ can be chosen to have the same form as the ones that enter the Hopf maps from the flat space-time. Also, we find a decomposition of the electric and magnetic fields similar to the Clebsch parametrization in the Minkowski space-time in which the fields are explicitly real. For completeness, we give the Einstein-Maxwell equations in terms of the energy, momentum and stress-tensor of the electromagnetic field in the field line representation in the Section 4. The last section is devoted to discussions. In the Appendix we have collected some basic mathematical relations the have been used in the calculations.

\section{Einstein-Maxwell equations}

In this section, we establish our notations and review the Einstein-Maxwell equations in the $1+3$ formalism following \cite{Gourgoulhon:2007ue}. 

Consider a hyperbolic four-dimensional space-time manifold 
$\mathcal{M}$ endowed with a metric tensor field $\mathbf{g}$ of signature $(-,+,+,+)$ and a global scalar field $\mathbf{t}$. Then $\mathcal{M} \simeq \mathbb{R}\times \Sigma$ is a foliation generated by $\mathbf{t}$ with the leaves defined as $\Sigma_t = \{ p \in \mathcal{M} : \mathbf{t}(p) = t = \mbox{constant} \}$. The normal vector $\mathbf{n}$, the normal evolution vector $\mathbf{m}$ and the lapse function $N$ are defined as 
\begin{equation}
\mathbf{n} := -N \nabla \mathbf{t},
\, \, \, \, \,
\mathbf{m} := N \mathbf{n},
\, \, \, \, \,
N := 
\left[
- g_{\mu \nu}\nabla^{\mu} \mathbf{t} \nabla^{\nu} \mathbf{t}
\right]^{-\frac{1}{2}},
\label{normal-normal-evolution-Lapse}
\end{equation}
where $\mu, \nu = 0, 1, 2, 3$ are space-time indices. It is further required that the gradient $\nabla \mathbf{t}$ be time-like which makes the leaves $\Sigma_t$ space-like. In the neighbourhood $U_p \in \Sigma_t $ we choose  the local coordinates $(t,x)=(t,x^i)$ adapted to the foliation, where $i,j = 1, 2, 3$ are the indices for the objects living on the leaf. Then the time evolution of any system is determined by the following vector
\begin{equation}
\pmb{\partial}_{t} := \mathbf{m} + \pmb{\beta},
\, \, \, \, \,
g_{\mu \nu} \beta^{\mu} n^{\nu} = 0,
\label{relations-adapted-coordinates}
\end{equation}
where $\pmb{\partial}_{t}$ is the derivative along the adapted time and $\pmb{\beta}$ is the shift vector from $\mathcal{T}_p (\mathcal{M})$. The induced metric on $\Sigma_t$ is denoted by 
$\pmb{\gamma}$ and has the components $\gamma_{ij} = g_{ij}$. 
This is obtained by projecting the tensor field $\mathbf{g}$ onto $\Sigma_t$ with the projector ${P^{\mu}}_{\nu} := {g^{\mu}}_{\nu} + n^{\mu}n_{\nu}$. The metric $\pmb{\gamma}$ is compatible with the covariant derivative $D_i$, i. e. $D^{i}\gamma_{ij} = 0$, which makes the associated connection a metric connection assumed to be torsionless. The components of the extrinsic curvature are defined as usual by the following relations
\begin{equation}
K_{ij}  : = - \pmb{\partial}^{\mu}_{i} \pmb{\partial}^{\nu}_{j}
\nabla_{\nu} n_{\mu},
\, \, \, \, \, 
K:= \gamma^{ij} K_{ij}.
\label{extrinsic-curvature}
\end{equation} 
The Maxwell field strength tensor $F_{\mu \nu}(t,x)$ can be decomposed locally in to the electric $E^{\mu} (t,x)$ and magnetic $B^{\mu}(t,x)$ components, respectively, as follows
\begin{equation}
E_{\mu}(t,x) = F_{\mu \nu}(t,x)n^{\nu} (t,x),
\hspace{0.5cm}
B_{\mu}(t,x) = \frac{1}{2} \varepsilon_{\mu \nu \sigma} (t,x)F^{\nu \sigma}(t,x),
\label{electric-magnetic-decomposition}
\end{equation}
where $\varepsilon_{\mu \nu \sigma}(t,x)$ is the contracted four-dimensional Levi-Civita tensor (see the equation (\refeq{Levi-Civita-4-3-indices})from the Appendix). The vectors $E^{\mu} (t,x)$ and $B^{\mu}(t,x)$ belong to $\mathcal{T}_p(\Sigma_t)$ due to the following relations
\begin{equation}
E_{\mu} (t,x) n^{\mu} (t,x) = 0,
\hspace{0.5cm}
B_{\mu}(t,x) n^{\mu} (t,x) = 0.
\label{electric-magnetic-fields-tangent}
\end{equation}
Then the Maxwell field strength tensor has the following form\footnote{In what follows we will drop the local space-time coordinates most of the time, unless the locality needs to be emphasized.}
\begin{equation}
F_{\mu \nu} = n_{\mu} E_{\nu} - n_{\nu} E_{\mu} + \varepsilon_{\mu \nu \rho \sigma} n^{\rho} B^{\sigma}.
\label{Electromagnetic-field-tensor}
\end{equation}

The dynamics of the electromagnetic field in the presence of gravity is given by the Maxwell's equations of the Einstein-Maxwell theory. In the absence of sources of the electromagnetic field and in the adapted coordinate frame, the Maxwell's equations take the following form
\begin{align}
\mathcal{L}_{\mathbf{m}} E^{i} - NKE^{i} - 
\varepsilon^{ijk} D_{j} \left( N B_{k} \right) & = 0,
\label{Faraday-gravitation}
\\
\mathcal{L}_{\mathbf{m}} B^{i} - NKB^{i} + 
\varepsilon^{ijk} D_{j} \left( N E_{k} \right) & = 0,
\label{Ampere-gravitation}
\\
D_{i}E^{i} &= 0,
\label{Gauss-E-gravitation}
\\
D_{ i} B^{i} &= 0.
\label{Gauss-B-gravitation}
\end{align}
As well known, the first two equations (\refeq{Faraday-gravitation}) and (\refeq{Ampere-gravitation}) determine the dynamics of the electric and magnetic fields, respectively. These are the generalization of the Faraday's and Amp\`{e}re's laws. On the other hand, the equations (\refeq{Gauss-E-gravitation}) and (\refeq{Gauss-B-gravitation}) generalize the Gauss' laws and represent constraints that $E^i$ and $B^i $ must obey at all times.

The definition of the Lie derivative $\mathcal{L}_{\mathbf{m}}$ and the three-dimensional Levi-Civita tensor are given by the equations (\refeq{Lie-derivative-properties}) and (\refeq{Levi-Civita-tensor}), respectively, from the Appendix. For other details we refer the reader to \cite{Gourgoulhon:2007ue}. 

\section{Existence of the general field line solutions}

In this section, we are going to show that there are local general solutions to the Maxwell's equations (\refeq{Faraday-gravitation}) - (\refeq{Gauss-B-gravitation}) that can be expressed in terms of the field lines of the vector fields $\mathbf{E}$ and $\mathbf{B}$ and generalize the corresponding solutions from the Minkowski space-time.

\subsection{Magnetic field line solutions}

Let us start with the equations (\refeq{Ampere-gravitation}) and (\refeq{Gauss-B-gravitation}). The magnetic field lines can be defined locally as follows. One picks up a neighbourhood $U_p \in \mathcal{M}$ where $p \in \Sigma_t$ and chooses the adapted coordinates $(t,x)$ in $U_p$. Then one can define a complex scalar field $\phi : U_p \rightarrow \mathbb{C}$ such that the field lines of the magnetic field $B^{i}(t,x)$ are the level curves of $\phi(t,x)$ in $U_p$. The vector field $B^{i}(t,x)$ must obey the constraint given by the equation (\refeq{Gauss-B-gravitation}) while its dynamics is given by the equations (\refeq{Ampere-gravitation}). The equation (\refeq{Gauss-B-gravitation}) suggests that the magnetic field should have the following form
\begin{equation}
B^{i} (t,x) = f(t,x) \varepsilon^{ijk} (t,x) D_{j} \phi (t,x) D_{k} \bar{\phi} (t,x),
\label{magnetic-field-ansatze}
\end{equation}
where the bar denotes the complex conjugate and $f(t,x)$ is an arbitrary real field on $U_p$. One can easily see that the above field $B^{i} (t,x)$ satisfies the constraint (\refeq{Gauss-B-gravitation}) if the function $f(t,x)$ depends on the coordinates implicitly as $f(\phi(t,x),\bar{\phi}(t,x))$ which is the same condition as in the flat space-time. 

From the equation (\refeq{Ampere-gravitation}), one can see that the time-evolution of the field $B^{i} (t,x)$ is determined by both the gravitational and the electric field $E^i (t,x)$. That suggests taking for the electric field the following expression 
\begin{equation}
E^i (t,x) = \frac{f(t,x)}{N(t,x)} 
\left[
\left( \mathcal{L}_{\mathbf{m}}  \bar{\phi} (t,x) \right) D^{i}\phi (t,x) 
-
\left( \mathcal{L}_{\mathbf{m}}  \phi (t,x) \right) D^{i} \bar{\phi} (t,x)
\right].
\label{electric-field-ansatze}
\end{equation}
In order to show that the fields $B^{i}(t,x)$ and $E^{i}(t,x)$ from the equations (\refeq{magnetic-field-ansatze}) and (\refeq{electric-field-ansatze}) satisfy the equation (\refeq{Ampere-gravitation}), 
we proceed at direct calculation of the terms from both sides of the Amp\`{e}re's law. Firstly, we expand the terms that contain $B^{i}(t,x)$. Then we note that some of the factors generated by the covariant derivative are symmetric while the Levi-Civita tensor is antisymmetric and thus many of these terms cancel individually. After some more algebraical manipulations, it turns out that the only non-vanishing terms from the right hand side of the equation (\refeq{Ampere-gravitation}) are the following ones 
\begin{equation}
+
\varepsilon^{ijk} f
D_j 
\left[
\left( \mathcal{L}_{\mathbf{m}} \bar{\phi} \, \right) D_{k}\phi  
-
\left( \mathcal{L}_{\mathbf{m}} \phi \right) D_{k}\bar{\phi} \,
\right].
\label{non-vanishing-lfs-Ampere-gravitation}
\end{equation}
The formula (\refeq{non-vanishing-lfs-Ampere-gravitation}) has been derived by using the basic relations (\refeq{definition-covariant-derivative})- (\refeq{Levi-Civita-tensor}) from the Appendix. 
The expression from the equation (\refeq{non-vanishing-lfs-Ampere-gravitation}) should be equated with the left hand side of the equation (\refeq{Ampere-gravitation}) with (\refeq{electric-field-ansatze}). After some algebra, one can show that the left hand side of the equation (\refeq{Ampere-gravitation}) takes the following form
\begin{equation}
- \varepsilon^{ijk} D_j 
\left[
f
\left(
\left(
\mathcal{L}_{\mathbf{m}}  \bar{\phi} \right) D_{k}\phi
-
\left( \mathcal{L}_{\mathbf{m}} \phi \right) D_{k}\bar{\phi} \
\right)
\right].
\label{non-vanishing-rfs-Ampere-gravitation}
\end{equation}
These terms can be expanded further by applying successively the covariant derivative (\refeq{definition-covariant-derivative}) and  the Lie derivative (\refeq{definition-Lie-derivative}). Then some straightforward calculations result in the equality of both terms from (\refeq{non-vanishing-lfs-Ampere-gravitation}) and (\refeq{non-vanishing-rfs-Ampere-gravitation}) which can be shown to have the same form
\begin{align}
& \varepsilon^{ijk} f 
\left(
\partial_t \partial_j \phi - \beta^r \partial_r \partial_j \phi - \partial_r \phi \partial_j \beta^r
\right)\partial_k \bar{\phi}
\nonumber
\\
& +
\varepsilon^{ijk} f 
\left(
\partial_t \partial_k \bar{\phi} - \beta^r \partial_r \partial_k \bar{\phi} - \partial_r \bar{\phi} \partial_k \beta^r
\right)\partial_j \phi.
\label{common-form-lhs-rhs-Ampere-gravitation}
\end{align}
That proves that the fields $B^{i}(t,x)$ and $E^{i}(t,x)$ from the equations (\refeq{magnetic-field-ansatze}) and (\refeq{electric-field-ansatze}) are field line solutions of the equations (\refeq{Ampere-gravitation}) and (\refeq{Gauss-B-gravitation}). We note that while $B^{i}(t,x)$ has a clear geometric interpretation in terms of magnetic lines, the field $E^i (t,x)$ does not and its connection with the electric lines is obscured in the equation (\refeq{electric-field-ansatze}). Indeed, this less clear geometrical picture is due to the fact that the electric and the magnetic fields are reciprocally orthogonal $\gamma^{ij}E_i B_j = 0$ at every point of $U_p$ as can be easily proved.

\subsection{Dual electric field line solutions}

In order to express the electric field in terms of its field lines, we firstly recall that the electric and magnetic fields away from sources are dual to each other in the curved space-time \cite{Deser:1976iy}. Therefore, one can look for solutions to the Gauss' law (\refeq{Gauss-E-gravitation}) for the electric field and the Faraday's law (\refeq{Faraday-gravitation}) by exploiting this duality as is done in the flat space-time. 

In the dual picture, the field lines of the electric field are the level curves of a second complex scalar field $\theta: U_p \rightarrow \mathbb{C}$. Since $E^i (t,x)$ obeys the same constraint equation as $B^{i}(t,x)$, it is natural to take it of the following form
\begin{equation}
E^{i} (t,x) = g(t,x) \varepsilon^{ijk} (t,x) D_{j} \bar{\theta} (t,x) D_{k} \theta (t,x),
\label{electric-field-ansatze-1}
\end{equation}
where $g(t,x)$ is an arbitrary real smooth field on $U_p$. The associated magnetic field $B^{i}(t,x)$ that determines the time-evolution of the field $E^i (t,x)$ is given by the following relation 
\begin{equation}
B^i (t,x) = \frac{g(t,x)}{N(t,x)} 
\left[
\left( \mathcal{L}_{\mathbf{m}}  \bar{\theta} (t,x) \right) D^{i}\theta (t,x) 
-
\left( \mathcal{L}_{\mathbf{m}}  \theta (t,x) \right) D^{i} \bar{\theta} (t,x)
\right].
\label{magnetic-field-ansatze-1}
\end{equation}
Following the same steps as before, one can show that the equation (\refeq{Gauss-E-gravitation}) is satisfied if $g(t,x)$ depends implicitly on the adapted coordinates $g(\theta (t,x),\bar{\theta}(t,x))$. Also, by reproducing the same calculations from the previous subsection, one can prove that the electric and magnetic fields from the relations (\refeq{electric-field-ansatze-1}) and (\refeq{magnetic-field-ansatze-1}) satisfy the second set of Maxwell's equations (\refeq{Gauss-E-gravitation}) and (\refeq{Faraday-gravitation}). 

The solutions of the two sets of Maxwell's equations (\refeq{Ampere-gravitation}) and (\refeq{Gauss-B-gravitation}) and (\refeq{Gauss-E-gravitation}) and (\refeq{Faraday-gravitation}) represent the same electromagnetic field but in two different representations that display different field lines. Therefore, the scalar fields $f(\phi,\bar{\phi})$ and $g(\theta,\bar{\theta})$ are not independent of each other. Indeed, by requiring that the two sets of Maxwell's equations (\refeq{Ampere-gravitation}) and (\refeq{Gauss-B-gravitation}) and (\refeq{Gauss-E-gravitation}) and (\refeq{Faraday-gravitation}), respectively, be satisfied simultaneously, the following set of constraints are imposed on the scalar fields
\begin{align}
f(\phi,\bar{\phi}) \varepsilon^{ijk} 
D_j \phi  D_k \bar{\phi} 
& = 
\frac{g(\theta,\bar{\theta})}{N} 
\left[
\left( \mathcal{L}_{\mathbf{m}}  \bar{\theta}  \right) D^{i}\theta  
-
\left( \mathcal{L}_{\mathbf{m}}  \theta \right) D^{i} \bar{\theta} 
\right],
\label{f-g-nonlinear-equations-1}
\\
g(\theta,\bar{\theta}) \varepsilon^{ijk}  D_{j} \bar{\theta}  D_{k} \theta 
& =
\frac{f(\phi,\bar{\phi})}{N} 
\left[
\left( \mathcal{L}_{\mathbf{m}}  \bar{\phi} \right) D^{i}\phi  
-
\left( \mathcal{L}_{\mathbf{m}}  \phi \right) D^{i} \bar{\phi} \right].
\label{f-g-nonlinear-equations-2}
\end{align}
The above equations are the equivalent of the corresponding mutual constraints of the real functions $f$ and $g$ in the flat space-time.  
As is the case there, the equations (\refeq{f-g-nonlinear-equations-1}) and (\refeq{f-g-nonlinear-equations-2}) 
are non-linear, but with a higher degree of non-linearity due to the interaction with the gravitational field. 

This concludes our proof of the existence of field line solutions to the Maxwell's equations on general hyperbolic space-times.

Some comments are in order here. Our approach to solving the Maxwell's equations in this section is the same as used in the flat space-time by Ra\~{n}ada in \cite{Ranada:1989wc,Ranada:1990}. The difference here is the presence of the gravitational field that manifests itself in the covariant and the Lie derivatives of the complex scalars as well as in the presence of the gauge variables $(N, \pmb{\beta})$. The benefits of maintaining the close analogy with the flat space-time are two fold: on one hand, the local field line solutions have a simple and intuitive geometrical interpretation. On the other hand, one can easily argue that the curved space-time solutions reduce to the corresponding  flat space-time solution in the appropriate limit.  Indeed, by recalling that the foliation of the Minkowski space-time is trivial, it follows that one can fix the gravitational gauge by picking up the Gauss normal coordinate systems ($N=1, \pmb{\beta} = 0$) in which our fields take the form of the solutions from \cite{Ranada:1989wc,Ranada:1990}. Thus, the fields obtained here are solutions of the Maxwell's equations in the field line representations in exactly the same way as in the flat space-time. That shows that our proposal is a local generalization of the field configurations given in \cite{Ranada:1989wc,Ranada:1990} to the Einstein-Maxwell theory. 

\subsection{Real and topological solutions}

We have proved in the previous subsection that the electromagnetic field is described by local field line solutions of the Maxwell's equations parametrized by the two smooth scalar fields $f$ and $g$ that are arbitrary up to the constraints given by the equations (\refeq{f-g-nonlinear-equations-1}) and (\refeq{f-g-nonlinear-equations-2}) as in the Minkowski space-time. 

One interesting solutions studied in the flat space-time is given by the following choice of parameters
\begin{align}
f &= \frac{1}{2 \pi i} \frac{1}{(1+|\phi|^2)^2},
\label{topological-f}
\\
g &= \frac{1}{2 \pi i} \frac{1}{(1+|\theta|^2)^2}.
\label{topological-g}
\end{align} 
We can use these scalars to construct the electric and magnetic fields in the curved space-time. Direct calculations show that $E^i (t,x)$ and $B^i (t,x)$ obtained from (\refeq{topological-f}) and (\refeq{topological-g}) have the following form
\begin{align}
B^i (t,x) & = \varepsilon^{ijk} D_j \alpha_1 (t,x) D_k \alpha_2 (t,x),
\label{magnetic-field-ansatze-2}
\\
E^i (t,x) & = \varepsilon^{ijk} D_j \beta_1 (t,x) D_k \beta_2 (t,x),
\label{electric-field-ansatze-2}
\end{align}
where $\alpha_1$, $\alpha_2$, $\beta_1$ and $\beta_2$ are real scalar fields expressed in the coordinates associated to the foliation in $U_p$. The equations (\refeq{magnetic-field-ansatze-2}) and (\refeq{electric-field-ansatze-2}) represent the same fields as the equations (\refeq{magnetic-field-ansatze}) and (\refeq{electric-field-ansatze-1}) if the following identifications are made
\begin{align}
\alpha_1 &= \frac{1}{1+|\phi|^2},
\hspace{0.5cm}
\alpha_2 = \frac{\Phi}{2 \pi},
\hspace{0.5cm}
\phi = |\phi| e^{i \Phi},
\label{alpha-phi-identification}
\\
\beta_1 &= \frac{1}{1+|\theta|^2},
\hspace{0.5cm}
\beta_2 = \frac{\Theta}{2 \pi},
\hspace{0.5cm}
\theta = |\theta| e^{i \Theta}.
\label{beta-phi-identification}
\end{align}
The proof that $E^i(t,x)$ and $B^i(t,x)$ can be expressed in terms of the fields $\alpha_{1,2}$ and $\beta_{1,2}$ from the equations (\refeq{alpha-phi-identification}) and (\refeq{beta-phi-identification}) above can be made by 
calculating explicitly the two expressions of $E^i(t,x)$ and $B^i(t,x)$ in terms of 
$\phi$, $\alpha_1$, $\alpha_2$ and $\theta$, $\beta_1$, $\beta_2$ , respectively. Also, we can show that the electric and magnetic fields obtained in this way are real. 

The solutions given by the equations (\refeq{magnetic-field-ansatze-2}) and (\refeq{electric-field-ansatze-2}) are interesting because they represent the local generalization of the electromagnetic solutions from the Minkowski space-time used to construct the electromagnetic Hopf knots \cite{Ranada:1989wc,Ranada:1990}. However, despite the fact that the local analysis of the Maxwell's equations leads to results similar to the ones obtained in the flat space-time, there are no topological (or even global) solutions on a general hyperbolic manifold as the existence of such solution depends on the topology of the leaf $\Sigma$.

An exception is the particular case $\Sigma \simeq S^3$ which corresponds to the cosmological models with the $SO(3)$ symmetry. Because $S^3$ is parallelizable, there are no topological obstructions to constructing non-vanishing global vector fields on it. Moreover, there is a global frame $\{ \omega^i \} = \{ \omega^1, \omega^2, \omega^3 \}$ that obeys the $su (2)$ algebra on $S^3$ and constitutes the basis of the cotangent space $\mathcal{T}^*(S^3)$ at each point. Since the global frame is not a global coordinate frame, one has to reformulate the Maxwell's equations in terms of forms. Then, by choosing the Gauss normal gauge $(N=1, \pmb{\beta}=0)$
and defining the time evolution in the holonomic basis of the cylinder, one obtains a Hopf knot field. This solution was given in \cite{Kopinski:2017nvp}.

\section{The Einstein's equations}

In the previous section, we have proved that $E^i(t,x)$ and $B^i(t,x)$ are field line solutions 
to the Einstein-Maxwell theory. For completeness, we give in this section the Einstein's equations for the fields given by the relations (\refeq{magnetic-field-ansatze}) and (\refeq{electric-field-ansatze}), respectively. By using them, one can readily calculate the trace of the energy density $\mathcal{E}$, the momentum density $p^{i}$ and the stress tensor $S_{ij}$ for the corresponding electromagnetic field. By plugging the results obtained in this way into the general form of the Einstein's equations in the $1+3$-formulation 
\cite{Gourgoulhon:2007ue} we obtain the following dynamical equations for the gravitational field
\begin{align}
\mathcal{L}_{\mathbf{m}} \gamma_{ij} & = -2NK_{ij},
\label{Einstein-eq-matric-dynamics}
\\
\mathcal{L}_{\mathbf{m}} K_{ij} & = - D_i D_j N + 
N \left( 
R_{ij} + K K_{ij} -2 K_{ir} {K^{r}}_j
\right)
\nonumber
\\
& -
\frac{4 \pi f^2}{N^2}
\left(
\left(\mathcal{L}_{\mathbf{m}} \bar{\phi}  \right) D_{k} \phi
-
\left(\mathcal{L}_{\mathbf{m}} \phi \right) D_{k} \bar{\phi} \,
\right)
\left(
\left(\mathcal{L}_{\mathbf{m}} \bar{\phi} \right) D^{k} \phi
-
\left(\mathcal{L}_{\mathbf{m}} \phi \right) D^{k} \bar{\phi} \,
\right) \gamma_{ij}
\nonumber
\\
& - \pi f^{2}
\left(
D^r \phi D^s \bar{\phi} - D^{s} \phi D^{r} \bar{\phi}
\right)
D_r \phi D_s \bar{\phi} \, \gamma_{ij}
\nonumber
\\
& + \frac{8 \pi f^2}{N^2}
\left(
\left(\mathcal{L}_{\mathbf{m}} \bar{\phi}  \right) D_{i} \phi
-
\left(\mathcal{L}_{\mathbf{m}} \phi \right) D_{i} \bar{\phi} \,
\right)
\left(
\left(\mathcal{L}_{\mathbf{m}} \bar{\phi} \right) D_{j} \phi
-
\left(\mathcal{L}_{\mathbf{m}} \phi \right) D_{j} \bar{\phi} \,
\right)
\nonumber
\\
& -
\varepsilon_{imn} \varepsilon_{jrs}
D^{m} \phi D^{n} \bar{\phi} D^{r} \phi D^{s} \bar{\phi}.
\label{Einstein-eq-curvature-dynamics}
\end{align}
In the same way, one can obtain the Hamiltonian and the momentum constraints that take the following form
\begin{align}
R + K^2 - K_{ij} K^{ij}
& = 
8 \pi f^2
\left[
\left(\mathcal{L}_{\mathbf{m}} \bar{\phi}  \right)^2 D_{i} \phi D^{i} \phi
- 2 
\left(\mathcal{L}_{\mathbf{m}} \bar{\phi} \right) 
\left(\mathcal{L}_{\mathbf{m}} \phi \right)
D_{i} \bar{\phi} D^{i} \phi
\right.
\nonumber
\\
& +
\left.
\left(\mathcal{L}_{\mathbf{m}} \phi  \right)^2 D_{i} \bar{\phi} D^{i} \bar{\phi}
 + D^i \phi D^j \bar{\phi}
\left( 
D_{i} \phi D_j{i} \bar{\phi} -
D_{i} \bar{\phi} D_{j} \phi
\right)
\right],
\label{Einstein-eq-hamiltonian-constraint}
\\
D_{j}{K^j}_i - D_i K & = \frac{8 \pi f^2}{N}
\left(\mathcal{L}_{\mathbf{m}} \bar{\phi}  D^j \phi 
- 
\mathcal{L}_{\mathbf{m}} \phi  D^j \bar{\phi} 
\right)
\left(
D_i \phi D_j \bar{\phi} -
D_j \phi D_i \bar{\phi}
\right).
\label{Einstein-eq-momentum-constraint}
\end{align}
Similar equations can be obtained for the dual and the two scalar solutions. These are the most general equations for the evolution of the gravitational field generated by the field line solutions of the Maxwell's equations. As usual, 
in order to solve the Einstein's equations one needs to make further assumptions about the space-time and by choosing a convenient gauge $(N,\pmb{\beta})$.

\section{Discussions}

In this paper, we have proved that the Maxwell's equations of the Einstein-Maxwell theory admit a general class of local field line solutions that are a natural generalization of the solutions from the Minkowski space-time given in \cite{Ranada:1989wc,Ranada:1990}. The generalization is not trivial, as it involves non-trivial relations between the electromagnetic and the gravitational fields. Note that, while in the flat space-time, these solutions can be used to construct global electromagnetic knots, the global solutions are not always possible on an arbitrary hyperbolic manifold. Nevertheless, at least one particular example of our general solution is known in the case of spherical leafs \cite{Kopinski:2017nvp}. However, it would be interesting to explore further the electromagnetic line configurations to determine other local as well as global solutions of the Maxwell's equations. This problem can be studied  by performing an analysis akin of ours. Also, it would be interesting to obtain solutions to the Einstein-Maxwell theory in particular cases as there could be potentially new applications in astrophysics and cosmology as well as in the fundamental study of the topological knots.

\section*{Acknowledgements}
It is a pleasure to acknowledge H. Nastase for correspondence and M. C. Rodriguez for discussions.

\section{Appendix}

In this Appendix we have collected from \cite{Gourgoulhon:2007ue} few mathematical relations that were used throughout the text.

We recall that the action of the covariant derivative and of the Lie derivative on an arbitrary tensor from $\mathcal{T}^{(r,s)}(\Sigma_p)$ are given by the following equations
\begin{align}
D_j {T^{i_{1}\ldots i_{r}}}_{k_{1}\ldots k_{s}}=
&\partial_j {T^{i_{1}\ldots i_{r}}}_{k_{1}\ldots k_{s}}
\nonumber
\\
&+\,{\Gamma ^{i_{1}}}_{mj}{T^{mi_{2}\ldots i_{r}}}_{k_{1}\ldots k_{s}}
+\cdots 
+{\Gamma ^{i_{r}}}_{mj}{T^{i_{1}\ldots i_{r-1}m}}_{k_{1}\ldots k_{s}}
\nonumber
\\
&-\,{\Gamma ^{m}}_{k_{1}j}{T^{i_{1}\ldots i_{r}}}_{mk_{2}\ldots k_{s}}
-\cdots 
-{\Gamma ^{m}}_{k_{s}j}{T^{i_{1}\ldots i_{r}}}_{k_{1}\ldots k_{s-1}m},
\label{definition-covariant-derivative}
\\
({\mathcal  {L}}_{\pmb{\xi}}T)^{{i_{1}\ldots i_{r}}}{}_{{k_{1}\ldots k_{s}}}
& =\xi^{j}(\partial _{j}T^{{i_{1}\ldots i_{r}}}{}_{{k_{1}\ldots k_{s}}})
\nonumber
\\
&-(\partial _{j}\xi^{{i_{1}}})T^{{ji_{2}\ldots i_{r}}}{}_{{k_{1}\ldots k_{s}}}
-\ldots 
-(\partial _{j}\xi^{{i_{r}}})T^{{i_{1}\ldots i_{{r-1}}j}}{}_{{k_{1}\ldots k_{s}}}
\nonumber
\\
&+(\partial _{{k_{1}}}\xi^{j})T^{{i_{1}\ldots i_{r}}}{}_{{jk_{2}\ldots k_{s}}}
+\ldots +(\partial _{{k_{s}}}\xi^{j})T^{{i_{1}\ldots i_{r}}}{}_{{k_{1}\ldots k_{{s-1}}j}},
\label{definition-Lie-derivative}
\end{align}
where $\pmb{\xi}$ is an arbitrary vector field from $\mathcal{T}(\Sigma)$. The Lie derivative of $\mathbf{m}$, $\pmb{\partial}_{\mathbf{t}}$ and $\pmb{\beta}$ satisfy the following relations
\begin{equation}
\mathcal{L}_{\mathbf{m}} : = \mathcal{L}_{\pmb{\partial}_{\mathbf{t}}} - \mathcal{L}_{\pmb{\beta}}
= \partial_{t} - \mathcal{L}_{\pmb{\beta}},
\label{Lie-derivative-properties}
\end{equation}
The second equality the equations (\refeq{Lie-derivative-properties}) holds in the adapted coordinates to the leaf while the first equality is valid for all tensor fields from $\mathcal{T}^{(r,s)}(\mathcal{M})$. In particular, the time evolution of the induced metric is given by the following equation
\begin{equation}
\mathcal{L}_{\mathbf{m}} = - 2 N K_{ij}.
\label{Einstein-equation-1}
\end{equation} 
It is important to recall that the Levi-Civita tensor on the leaf satisfies the relations
\begin{equation}
\varepsilon_{ijk} : = \sqrt{\gamma} \epsilon_{ijk},
\hspace{0.5cm}
\gamma^{ij}D_{i}\varepsilon_{jrs} = 0,
\label{Levi-Civita-tensor}
\end{equation}
where $\sqrt{\gamma} = \det \gamma_{ij}$. The four-dimensional Levi-Civita tensor $\varepsilon_{\mu \nu \rho \sigma}$ is the four-dimensional volume element. It can be contracted with the normal vector $n^{\mu}$ to obtain the three-indices four dimensional Levi-Civita tensor
\begin{equation}
\varepsilon_{\mu \nu \sigma} := 
n^{\rho} \varepsilon_{\rho \mu \nu \sigma},
\hspace{0.5cm}
\varepsilon_{\mu \nu \sigma} n^{\sigma} =0.
\label{Levi-Civita-4-3-indices}
\end{equation}


\begin{thebibliography}{Proper}


\bibitem{Trautman:1977im} 
  A.~Trautman,
  Int.\ J.\ Theor.\ Phys.\  {\bf 16}, 561 (1977).
  \doi{10.1007/BF01811088}

\bibitem{Ranada:1989wc} 
  A.~F.~Ra\~{n}ada,
  Lett.\ Math.\ Phys.\  {\bf 18}, 97 (1989).
  \doi{10.1007/BF00401864}

\bibitem{Ranada:1990}
A.~F.~Ra\~{n}ada,
  J.\ Phys.\ A {\bf 23}L815, (1990). 

\bibitem{Ranada:1992hw} 
  A.~F.~Ra\~{n}ada,
  J.\ Phys.\ A {\bf 25}, 1621 (1992).

\bibitem{Bateman:1915}
H.~ Bateman, 
``The Mathematical Analysis of Electrical and Optical Wave-motion on the Basis of Maxwell's Equations,
Cambridge University Press (1915).

\bibitem{Ranada:1995}
  A.~F.~Ra\~{n}ada, J.~L.~Trueba,
  Phys.\ Lett.\ A {\bf 202}, 337-342 (1995).
  \doi{10.1016/0375-9601(95)00352-4}

\bibitem{Ranada:1997}
  A.~F.~Ra\~{n}ada and J.~L.~Trueba,
  Phys.\ Lett.\ A {\bf 222}, 25-33 (1997).
  \doi{10.1016/S0375-9601(97)00366-6}

 

\bibitem{Irvine:2008}
  W.~T.~M.~Irvine, D.~Bouwmeester,
  Nature Physics 4, 716 - 720 (2008).
  \doi{10.1038/nphys1056} 

\bibitem{Irvine:2010}
  W.~T.~M.~Irvine,
  J.\ Phys.\ A:\ Math.\ Theor.\ {\bf 43} 385203 (2010).
  \doi{10.1088/1751-8113/43/38/385203}


\bibitem{Arrayas:2010xi} 
  M.~Array\'{a}s and J.~L.~Trueba,
  J.\ Phys.\ A {\bf 43}, 235401 (2010)
  \doi{10.1088/1751-8113/43/23/235401}
  [arXiv:1001.4985 [math-ph]].


\bibitem{Kleckner:2013}
  D.~ Kleckner and W.~ T.~ M.~ Irvine, 
  Nature Physics {\bf 9} 253-258 (2013).
  \doi{10.1038/nphys2560}

\bibitem{Arrayas:2011ci} 
  M.~Array\'{a}s and J.~L.~Trueba,
  Annalen Phys.\  {\bf 524}, 71 (2012)
  \doi{10.1002/andp.201100119}
  [arXiv:1105.6285 [hep-th]].

\bibitem{Arrayas:2017wtr} 
  M.~Array\'{a}s and J.~L.~Trueba,
  J.\ Phys.\ A {\bf 50}, no. 8, 085203 (2017)
  \doi{10.1088/1751-8121/aa54dd}
  [arXiv:1610.06014 [physics.class-ph]].
	
\bibitem{Ranada:2017ore} 
  A.~F.~Ra\~{n}ada, A.~Tiemblo and J.~L.~Trueba,
  Int.\ J.\ Geom.\ Meth.\ Mod.\ Phys.\  {\bf 14}, no. 05, 1750073 (2017).
  \doi{10.1142/S0219887817500736}


\bibitem{Ranada:1998vp} 
  A.~F.~Ra\~{n}ada and J.~L.~Trueba,
  Phys.\ Lett.\ B {\bf 422}, 196 (1998)
  \doi{10.1016/S0370-2693(98)00071-9}
  [hep-th/9802166].

\bibitem{Ranada:2003}
A.~ F.~ Ra\~{n}ada, 
  Phy.\ Lett.\ A {\bf 310}, 434-444(2003). 
  \doi{10.1016/S0375-9601(03)00443-2}

\bibitem{Ranada:2006tq} 
  A.~F.~Ra\~{n}ada and J.~L.~Trueba,
  Found.\ Phys.\  {\bf 36}, 427 (2006).
  \doi{10.1007/s10701-005-9026-8}

\bibitem{Arrayas:2012eja} 
  M.~Array\'{a}s, J.~L.~Trueba and A.~F.~Ra\~{n}ada,
  in Trends in Electromagnetism - From Fundamentals to Applications,
  V. Barsan and R. P. Lungu (Eds.), InTech (2012).
  \doi{10.5772/34703}


\bibitem{Arrayas:2011ia} 
  M.~Array\'{a}s and J.~L.~Trueba,
  2015 J. Phys. A: Math. Theor. 48 025203
  \doi{10.1088/1751-8113/48/2/025203}
  [arXiv:1106.1122 [hep-th]].

\bibitem{Kedia:2013bw} 
  H.~Kedia, I.~Bialynicki-Birula, D.~Peralta-Salas and W.~T.~M.~Irvine,
  Phys.\ Rev.\ Lett.\  {\bf 111}, 150404 (2013)
  \doi{10.1103/PhysRevLett.111.150404}
  [arXiv:1302.0342 [math-ph]].

\bibitem{Hoyos:2015bxa} 
  C.~Hoyos, N.~Sircar and J.~Sonnenschein,
  J.\ Phys.\ A {\bf 48}, no. 25, 255204 (2015)
  \doi{10.1088/1751-8113/48/25/255204}
  [arXiv:1502.01382 [hep-th]].


\bibitem{Goulart:2016orx} 
  E.~Goulart,
  Europhys.\ Lett.\  {\bf 115}, no. 1, 10004 (2016)
  \doi{10.1209/0295-5075/115/10004}
  [arXiv:1602.05071 [gr-qc]].

\bibitem{Alves:2017ggb} 
  D.~F.~W.~Alves, C.~Hoyos, H.~Nastase and J.~Sonnenschein,
  arXiv:1705.06750 [hep-th].

\bibitem{Alves:2017zjt} 
  D.~W.~F.~Alves, C.~Hoyos, H.~Nastase and J.~Sonnenschein,
  arXiv:1707.08578 [hep-th].


\bibitem{Ranada:1996}
  A.~F.~Ra\~{n}ada, J.~L.~Trueba,
  Nature {\bf 383}, 32 (1996).
  \doi{10.1038/383032a0}

\bibitem{Irvine:2014}
  W.~ T.~ M.~ Irvine, D.~ Kleckner,
  Nature Materials {\bf 13}, 229–231 (2014). 
  \doi{10.1038/nmat389}6

\bibitem{Smiet:2015}
  C.~ B.~ Smiet, S.~ Candelaresi, A.~ Thompson, J.~ Swearngin, J.~ W.~ Dalhuisen, D.~ Bouwmeester,
  Phys.\ Rev.\ Lett.\ {\bf 115}, 095001 (2015).
  \doi{10.1103/PhysRevLett.115.095001} 



\bibitem{Ren:2008zzf} 
  J.~R.~Ren, T.~Zhu and S.~F.~Mo,
  Commun.\ Theor.\ Phys.\  {\bf 50}, 1071 (2008).
  \doi{10.1088/0253-6102/50/5/12}

\bibitem{deKlerk:2017qvq} 
  A.~J.~J.~M.~de Klerk, R.~I.~van der Veen, J.~W.~Dalhuisen and D.~Bouwmeester,
  Phys.\ Rev.\ A {\bf 95}, no. 5, 053820 (2017).
  \doi{10.1103/PhysRevA.95.053820}
  

\bibitem{Trueba:2008sc}
J.~ L.~ Trueba, 
Annales de la Fondation Louis de Broglie, Vol. 33, no 1-2, (2008),
183-192.


\bibitem{Arrayas:2017sfq} 
  M.~Array\'{a}s, D.~Bouwmeester and J.~L.~Trueba,
  Phys.\ Rept.\  {\bf 667}, 1 (2017).
  \doi{10.1016/j.physrep.2016.11.001}



\bibitem{Dalhuisen:2012zz} 
  J.~W.~Dalhuisen and D.~Bouwmeester,
  J.\ Phys.\ A {\bf 45}, 135201 (2012).
  \doi{10.1088/1751-8113/45/13/135201}

\bibitem{Swearngin:2013sks} 
  J.~Swearngin, A.~Thompson, A.~Wickes, J.~W.~Dalhuisen and D.~Bouwmeester,
  arXiv:1302.1431 [gr-qc].

\bibitem{Thompson:2014pta} 
  A.~Thompson, J.~Swearngin and D.~Bouwmeester,
  J.\ Phys.\ A {\bf 47}, 355205 (2014)
  \doi{10.1088/1751-8113/47/35/355205}
  [arXiv:1402.3806 [gr-qc]].

\bibitem{Thompson:2014owa} 
  A.~Thompson, A.~Wickes, J.~Swearngin and D.~Bouwmeester,
  J.\ Phys.\ A {\bf 48}, no. 20, 205202 (2015)
  \doi{10.1088/1751-8113/48/20/205202}
  [arXiv:1411.2073 [gr-qc]].

\bibitem{Kopinski:2017nvp} 
  J.~Kopiński and J.~Natário,
  Gen.\ Rel.\ Grav.\  {\bf 49}, no. 6, 81 (2017)
  \doi{10.1007/s10714-017-2242-7}
  [arXiv:1702.04923 [gr-qc]].

\bibitem{Gourgoulhon:2007ue} 
  E.~Gourgoulhon,
  ``3+1 formalism and bases of numerical relativity,''
  gr-qc/0703035.


\bibitem{Deser:1976iy} 
  S.~Deser and C.~Teitelboim,
  Phys.\ Rev.\ D {\bf 13}, 1592 (1976).
  \doi{10.1103/PhysRevD.13.1592}

\end{thebibliography}
\end{document}